\documentclass[aip,rsi,preprint,groupedaddress,showmsc,superscriptaddress,floatfix]{revtex4}
\usepackage{amsmath,amsfonts,amssymb,caption,hyperref,color,epsfig,graphics,graphicx,latexsym,mathrsfs,revsymb,theorem,url,verbatim,epstopdf}

\usepackage[center]{titlesec}
\titleformat{\section}[hang]{\Large\bfseries\filcenter}{}{1em}{}
\titleformat{\subsection}[hang]{\bfseries}{}{1em}{}
\usepackage{amsmath,amsfonts,amssymb,caption,hyperref,color,epsfig,graphics,graphicx,latexsym,mathrsfs,revsymb,theorem,url,verbatim,epstopdf,cleveref}
\hypersetup{colorlinks,linkcolor={blue},citecolor={blue},urlcolor={red}}

\newtheorem{definition}{Definition}

\newtheorem{lemma}[definition]{Lemma}

\newtheorem{theorem}[definition]{Theorem}

\def\squareforqed{\hbox{\rlap{$\sqcap$}$\sqcup$}}
\def\qed{\ifmmode\squareforqed\else{\unskip\nobreak\hfil
\penalty50\hskip1em\null\nobreak\hfil\squareforqed
\parfillskip=0pt\finalhyphendemerits=0\endgraf}\fi}
\def\endenv{\ifmmode\;\else{\unskip\nobreak\hfil
\penalty50\hskip1em\null\nobreak\hfil\;
\parfillskip=0pt\finalhyphendemerits=0\endgraf}\fi}
\newenvironment{proof}{\noindent \textbf{{Proof.~} }}{\qed}
\def\Dbar{\leavevmode\lower.6ex\hbox to 0pt
{\hskip-.23ex\accent"16\hss}D}
\def\bpf{\begin{proof}}
\def\epf{\end{proof}}

\newcommand{\abs}[1]{\left\lvert {#1} \right\rvert}

\newcommand{\red}{\textcolor{red}}

\newcommand{\tbc}{\red{TO BE CONTINUED}.~}

\newcommand{\nc}{\newcommand}

\def\bea{\begin{eqnarray}}
\def\eea{\end{eqnarray}}
\def\beq{\begin{equation}}
\def\eeq{\end{equation}}
\def\bal{\begin{aligned}}
\def\eal{\end{aligned}}
\def\bma{\begin{bmatrix}}
\def\ema{\end{bmatrix}}

\def\z{\zeta}

\def\t{\theta}

\def\p{\pi}

\def\o{\omega}

\nc{\bbA}{\mathbb{A}} \nc{\bbB}{\mathbb{B}} \nc{\bbC}{\mathbb{C}}
\nc{\bbD}{\mathbb{D}} \nc{\bbE}{\mathbb{E}} \nc{\bbF}{\mathbb{F}}
\nc{\bbG}{\mathbb{G}} \nc{\bbH}{\mathbb{H}} \nc{\bbI}{\mathbb{I}}
\nc{\bbJ}{\mathbb{J}} \nc{\bbK}{\mathbb{K}} \nc{\bbL}{\mathbb{L}}
\nc{\bbM}{\mathbb{M}} \nc{\bbN}{\mathbb{N}} \nc{\bbO}{\mathbb{O}}
\nc{\bbP}{\mathbb{P}} \nc{\bbQ}{\mathbb{Q}} \nc{\bbR}{\mathbb{R}}
\nc{\bbS}{\mathbb{S}} \nc{\bbT}{\mathbb{T}} \nc{\bbU}{\mathbb{U}}
\nc{\bbV}{\mathbb{V}} \nc{\bbW}{\mathbb{W}} \nc{\bbX}{\mathbb{X}}
\nc{\bbZ}{\mathbb{Z}}

\nc{\bA}{{\bf A}} \nc{\bB}{{\bf B}} \nc{\bC}{{\bf C}}
\nc{\bD}{{\bf D}} \nc{\bE}{{\bf E}} \nc{\bF}{{\bf F}}
\nc{\bG}{{\bf G}} \nc{\bH}{{\bf H}} \nc{\bI}{{\bf I}}
\nc{\bJ}{{\bf J}} \nc{\bK}{{\bf K}} \nc{\bL}{{\bf L}}
\nc{\bM}{{\bf M}} \nc{\bN}{{\bf N}} \nc{\bO}{{\bf O}}
\nc{\bP}{{\bf P}} \nc{\bQ}{{\bf Q}} \nc{\bR}{{\bf R}}
\nc{\bS}{{\bf S}} \nc{\bT}{{\bf T}} \nc{\bU}{{\bf U}}
\nc{\bV}{{\bf V}} \nc{\bW}{{\bf W}} \nc{\bX}{{\bf X}}
\nc{\bZ}{{\bf Z}}

\nc{\bmA}{{\bm A}} \nc{\bmB}{{\bm B}} \nc{\bmC}{{\bm C}}
\nc{\bmD}{{\bm D}} \nc{\bmE}{{\bm E}} \nc{\bmF}{{\bm F}}
\nc{\bmG}{{\bm G}} \nc{\bmH}{{\bm H}} \nc{\bmI}{{\bm I}}
\nc{\bmJ}{{\bm J}} \nc{\bmK}{{\bm K}} \nc{\bmL}{{\bm L}}
\nc{\bmM}{{\bm M}} \nc{\bmN}{{\bm N}} \nc{\bmO}{{\bm O}}
\nc{\bmP}{{\bm P}} \nc{\bmQ}{{\bm Q}} \nc{\bmR}{{\bm R}}
\nc{\bmS}{{\bm S}} \nc{\bmT}{{\bm T}} \nc{\bmU}{{\bm U}}
\nc{\bmV}{{\bm V}} \nc{\bmW}{{\bm W}} \nc{\bmX}{{\bm X}}
\nc{\bmZ}{{\bm Z}}

\nc{\cA}{{\cal A}} \nc{\cB}{{\cal B}} \nc{\cC}{{\cal C}}
\nc{\cD}{{\cal D}} \nc{\cE}{{\cal E}} \nc{\cF}{{\cal F}}
\nc{\cG}{{\cal G}} \nc{\cH}{{\cal H}} \nc{\cI}{{\cal I}}
\nc{\cJ}{{\cal J}} \nc{\cK}{{\cal K}} \nc{\cL}{{\cal L}}
\nc{\cM}{{\cal M}} \nc{\cN}{{\cal N}} \nc{\cO}{{\cal O}}
\nc{\cP}{{\cal P}} \nc{\cQ}{{\cal Q}} \nc{\cR}{{\cal R}}
\nc{\cS}{{\cal S}} \nc{\cT}{{\cal T}} \nc{\cU}{{\cal U}}
\nc{\cV}{{\cal V}} \nc{\cW}{{\cal W}} \nc{\cX}{{\cal X}}
\nc{\cZ}{{\cal Z}}

\begin{document}
\title{Some special complex Hadamard matrices of order six}

\author{Mengfan Liang}\email[]{lmf2021@buaa.edu.cn}
\affiliation{LMIB(Beihang University), Ministry of Education, and School of Mathematical Sciences, Beihang University, Beijing 100191, China}

\author{Lin Chen}\email[]{linchen@buaa.edu.cn (corresponding author)}
\affiliation{LMIB(Beihang University), Ministry of Education, and School of Mathematical Sciences, Beihang University, Beijing 100191, China}
\affiliation{International Research Institute for Multidisciplinary Science, Beihang University, Beijing 100191, China}

\author{Fengyue Long}\email[]{lfy2021@buaa.edu.cn (corresponding author)}
\affiliation{LMIB(Beihang University), Ministry of Education, and School of Mathematical Sciences, Beihang University, Beijing 100191, China}

\author{Xinyu Qiu}\email[]{xinyuqiu@buaa.edu.cn (corresponding author)}
\affiliation{LMIB(Beihang University), Ministry of Education, and School of Mathematical Sciences, Beihang University, Beijing 100191, China}

\begin{abstract}
The complete classification of $6\times 6$ complex Hadamard matrices (CHMs) is a long-standing open problem. In this paper we investigate a series of CHMs, such as the CHMs containing a $2\times 3$ submatrix with rank one, the CHMs containing exactly three distinct elements and all elements of the first row being one, the $H_2$-reducible matrices containing exactly three distinct matrix elements, and the CHMs containing a $3\times 3$ Hadamard submatrix. We characterize all forms of these CHMs. Our results are the latest progress on the complete classification of CHMs. 
\end{abstract}

\date{\today}

\maketitle


\tableofcontents
\newpage

\section{Introduction}
\label{sec:intro}

For describing observables in quantum physics, Schwinger defined the MUBs \cite{schwinger60} in 1960. It is known that two quantum states in $\mathbb{C}^d$ are mutually unbiased (MU) when their inner product has modulus $\frac{1}{\sqrt d}$. Two orthogonal bases in $\mathbb{C}^d$ are MU when their elements are all MU. We refer to the complete MUBs as $d+1$ orthogonal bases in $\mathbb{C}^d$ and any two of them are MU. Ref.\cite{WOOTTERS1989363} found that the complete MUBs in dimension $d$ exist if $d$ is a prime power. So the first unknown case occurs at $d=6$. People have studied this case by many methods, such as the average distance between four bases in dimension six \cite{rle11}, Fourier family of Hadamard matrices \cite{jmm09},  computer program \cite{bw09},  noise robustness \cite{Designolle2018Quantifying},  product basis \cite{mw12jpa135307,mw12jpa102001,mpw16,Chen2018Mutually}  and so on   \cite{Boykin05,bw08,bw10,deb10,wpz11,mw12ijqi,Chen2017Product}. This case is also a long-standing open problem.
\par
We refer to the complex Hadamard matrix (CHM) H as an $n\times n$ matrix with elements of modulus one, and $\dfrac{1}{\sqrt n}$H is a unitary matrix. For convenience, we ignore the coefficient $\dfrac{1}{\sqrt n}$ of the CHM when we discuss four mutually unbiased bases (MUBs) containing the identity matrix in the following content. In this  
paper we investigate some $6\times 6$ CHMs. We prove that all $H_2$-reducible matrices defined by Karlsson in  \cite{karlsson11} containing exactly three distinct matrix elements are complex equivalent to the specific matrix $M_2$ in (\ref{le:eqn eight}). This is presented in Lemma \ref{le:three kinds of matrix elements and H2}. We find a new family of CHMs and we obtain that all $6\times 6$ CHMs containing a $3\times 3$ Hadamard submatrix are complex equivalent to the matrices from the new CHM family. They are the latest results on $6\times 6$ CHMs and they 
might lead to the complete classification of $6\times 6$ CHMs.
\tbc
\par
The rest of this paper is structured as follows. In Sec. \ref{sec:pre} we introduce some linear algebra, some properties of CHMs, as well as the parametrization of $H_2$-reducible matrices. In Sec. \ref{sec:H2} and \ref{sec:CHMs 33} we introduce the two main results of this paper. We have a complete classification with the $H_2$-reducible matrices containing exactly three distinct matrix elements and the CHMs containing a $3\times 3$ Hadamard submatrix. We conclude in Sec. \ref{sec:con}.

\section{Preliminaries}
	\label{sec:pre}

In this section we introduce the fundamental notations and facts used in this paper. We review some linear algebra in Sec. \ref{subsec:la}. Then we introduce a series of lemmas on CHMs in Sec. \ref{subsec:chm} from Lemma \ref{le:rank one} to \ref{le:two elements}. They will be used in the proof from Lemmas \ref{le:three kinds of matrix elements and H2} to \ref{le:33matrix}. In Sec. \ref{subsec:h2 and mub}, we review the definition and some properties of the $H_2$-reducible matrix.

\subsection{Linear algebra}
\label{subsec:la}
In this section we introduce some facts in linear algebra. 
\begin{lemma}
\label{le:abcd}	
(i) Suppose $a+b+c=0$ with complex numbers $a,b,c$ of modulus one. Then $(a,b,c)\propto(1,\o,\o^2)$ or $(1,\o^2,\o)$ where $\o=e^{2\p i\over 3}$.

(ii) Suppose $a+b+c+d=0$ with complex numbers $a,b,c,d$ of modulus one. Then $a=-b,-c$ or $-d$.

(iii) Suppose $g_1+g_2+g_3+k(g_4+g_5+g_6)=0$ with complex numbers $g_1,g_2,...,g_6$ of modulus one and $g_1,g_2,...,g_6\in \{1,\o ,\o^2 \}$ where $\o=e^{2\p i\over 3}$. If two of $g_1,g_2,g_3$ are not identical and $g_1+g_2+g_3\neq 0$, then $k\in \{1,\o ,\o^2 ,-1,-\o ,-\o^2 \}$.
\end{lemma}

\begin{proof}
Assertion $(i)$ and $(ii)$ are from the Lemma $2.5$ of \cite{lma2020}.
So we only prove $(iii)$. Without loss of generality, we assume that $g_1\neq g_2$. Then $g_1+g_2\in \{-1,-\o ,-\o^2 \}$. If $g_4=g_5=g_6$, then we have
$$(g_1+g_2)+g_3+3kg_4=0 \Longrightarrow 0=(kg_4)^*(g_1+g_2)+(kg_4)^*g_3+3\geq 3-1-1=1.
$$
It is a contradiction. So two of $g_4,g_5,g_6$ are not identical. Without loss of generality, we assume that $g_4\neq g_5$. Then $g_4+g_5\in \{-1,-\o ,-\o^2 \}$. Hence $(g_1+g_2)+g_3+k(g_4+g_5)+kg_6=0$. Using $(ii)$ we get $g_3=-k(g_4+g_5)$ or $g_3=-kg_6$. So $k\in \{1,\o ,\o^2 ,-1,-\o ,-\o^2 \}$ and we complete this proof.
\end{proof}

\subsection{CHM}
\label{subsec:chm}

To find out the relation between different CHMs, we define the complex equivalence. We refer to the \textit{monomial unitary matrix} as a unitary matrix each of whose rows and columns has exactly one nonzero entry. We say that two matrices $U$ and $V$ are complex equivalent when $U=PVQ$ where $P,Q$ are monomial unitary matrices. Specially when  $P,Q$ are both permutation matrices, we say that $U$ and $V$ are equivalent. When  $P,Q$ are both real monomial unitary matrices, we say that $U$ and $V$ are real equivalent. One can easily verify that if $U$ and $V$ are equivalent or real equivalent then $U$ and $V$ are complex equivalent. 
\par
The Tao matrix
\begin{eqnarray}\label{le:eqn ten}
S_6^{(0)}=\begin{bmatrix}
1 & 1 & 1 & 1 & s & s\\
1 & 1 & \o & \o & \o^2 & \o^2\\
1 & \o & 1 & \o^2 & \o^2 & \o \\
1 & \o & \o^2 & 1 & \o & \o^2 \\
1 & \o^2 & \o^2 & \o & 1 & \o \\
1 & \o^2 & \o & \o^2 & \o & 1
\end{bmatrix}
\end{eqnarray}
is a special CHM, and it is defined by $Terence\ Tao$ in \cite{Taom}, it plays an important role in the investigation of spectral sets. As far as we know, the Tao matrix is the only CHM in $6\times 6$ CHMs which does not belongs to the $H_2$-reducible matrices. 
\par
Next we are going to introduce two lemmas on CHMs.

\begin{lemma}
\label{le:rank one}	
If a CHM $X$ contains a $2\times 3$ submatrix with rank one, then $X$ is complex equivalent to the matrix from the following two-parameter family
\begin{eqnarray}\label{le:eqn one}	
H(\alpha ,\beta )=\begin{bmatrix}
1 & 1 & 1 & 1 & 1 & 1\\
1 & 1 & 1 &-1 & -1 & -1\\
1 & \o & \o^2 & \alpha  & \alpha \o  & \alpha \o^2 \\
1 & \o & \o^2 & -\alpha  & -\alpha \o & -\alpha \o^2 \\
1 & \o^2 & \o & \beta & \beta \o^2 & \beta \o \\
1 & \o^2 & \o & -\beta & -\beta \o^2 & -\beta \o
\end{bmatrix},
\end{eqnarray}
where $\o=e^{2\p i\over 3}$ and the parameters $\alpha$ and $\beta$ are both complex numbers of modulus one.
\end{lemma}
\begin{proof}
Suppose $X$ contains a $2\times 3$ submatrix with rank one, then $X$ is complex equivalent to 
\begin{eqnarray}
\label{le:eqn two}	
X^{'}=\begin{bmatrix}
1 & 1 & 1 & 1 & 1 & 1\\
1 & 1 & 1 & r_1 & r_2 & r_3\\
1 & a_1 & a_2 & a_3 & a_4 & a_5 \\
1 & b_1 & b_2 & b_3 & b_4 & b_5 \\
1 & c_1 & c_2 & c_3 & c_4 & c_5 \\
1 & d_1 & d_2 & d_3 & d_4 & d_5
\end{bmatrix},
\end{eqnarray}
where all elements in $X^{'}$ are complex numbers of modulus one. From (\ref{le:eqn two}) we have
\begin{eqnarray}
\label{le:eqn three}	
1+1+1+r_1+r_2+r_3=0 \Longrightarrow r_1=r_2=r_3=-1.
\end{eqnarray}
So by the orthogonality of the first three row vectors of $X^{'}$ we have
\begin{eqnarray}
\label{le:eqn four}	
\left\{
\begin{aligned}
1+a_1+a_2+a_3+a_4+a_5=0  \\
1+a_1+a_2-a_3-a_4-a_5=0 
\end{aligned}
\right.
\end{eqnarray}
Then we could solve (\ref{le:eqn four}) and get 
\begin{eqnarray}
\label{le:eqn five}	
1+a_1+a_2=a_3+a_4+a_5=0.
\end{eqnarray}
Using Lemma \ref{le:abcd} $(i)$ we obtain that $\{a_1,a_2 \}=\{\o,\o^2 \}$ and $(a_3,a_4,a_5)\propto(1,\o,\o^2)$ or $(1,\o^2,\o)$. 
Similarly we could get
$$\{b_1,b_2 \}=\{\o,\o^2 \},$$
$$\{c_1,c_2 \}=\{\o,\o^2 \},$$ 
$$\{d_1,d_2 \}=\{\o,\o^2 \},$$ and 
$$(b_3,b_4,b_5)\propto(1,\o,\o^2) \ or \  (1,\o^2,\o),$$ 
$$(c_3,c_4,c_5)\propto(1,\o,\o^2) \ or \ (1,\o^2,\o),$$ 
$$(d_3,d_4,d_5)\propto(1,\o,\o^2) \ or \ (1,\o^2,\o).$$
Obviously the CHM containing a $3\times 3$ submatrix with rank one does not exist. So it is easy to show that $X^{'}$ is complex equivalent to 
\begin{eqnarray}
\label{le:eqn six}	
X^{''}=\begin{bmatrix}
1 & 1 & 1 & 1 & 1 & 1\\
1 & 1 & 1 & -1 & -1 & -1\\
1 & \o & \o^2 & \alpha  & \alpha \o  & \alpha \o^2 \\
1 & \o & \o^2 & f_1 & f_2 & f_3 \\
1 & \o^2 & \o & f_4 & f_5 & f_6 \\
1 & \o^2 & \o & f_7 & f_8 & f_9
\end{bmatrix},
\end{eqnarray}
where all elements in $X^{''}$ are complex numbers of modulus one. Then using the orthogonality of the third and fourth row vectors of $X^{''}$ we could obtain that
\begin{eqnarray}
1+\o \o^*+\o^2 (\o^2)^*+\alpha f_1^*+\alpha \o f_2^*+\alpha \o^2 f_3^*=0.
\end{eqnarray}
Then
\begin{eqnarray}\label{le:eqn seven}	
X^{''}=\begin{bmatrix}
1 & 1 & 1 & 1 & 1 & 1\\
1 & 1 & 1 &-1 & -1 & -1\\
1 & \o & \o^2 & \alpha  & \alpha \o  & \alpha \o^2 \\
1 & \o & \o^2 & -\alpha  & -\alpha \o & -\alpha \o^2 \\
1 & \o^2 & \o & f_4 & f_5 & f_6 \\
1 & \o^2 & \o & f_7 & f_8 & f_9
\end{bmatrix}.
\end{eqnarray}
Using the orthogonality of the first, third and fifth row vectors of $X^{''}$ we could obtain that
\begin{eqnarray}
\left\{
\begin{array}{lr}
1+\o+\o^2+f_4+f_5+f_6 =0  \\
1+\o (\o^2)^*+\o^2 \o^*+\alpha f_4^*+\alpha \o f_5^*+\alpha \o^2 f_6^* =0 
\end{array}
\right.
\end{eqnarray}
Then
\begin{eqnarray}\label{le:eqn seven}	
X^{''}=\begin{bmatrix}
1 & 1 & 1 & 1 & 1 & 1\\
1 & 1 & 1 &-1 & -1 & -1\\
1 & \o & \o^2 & \alpha  & \alpha \o  & \alpha \o^2 \\
1 & \o & \o^2 & -\alpha  & -\alpha \o & -\alpha \o^2 \\
1 & \o^2 & \o & \beta & \beta \o^2 & \beta \o \\
1 & \o^2 & \o &  f_7 & f_8 & f_9
\end{bmatrix}.
\end{eqnarray}
Using the orthogonality of the fifth and sixth row vectors of $X^{''}$ we could obtain that
\begin{eqnarray}
1+\o^2 (\o^2)^*+\o \o^*+\beta f_7^*+\beta \o f_8^*+\beta \o^2 f_9^*=0.
\end{eqnarray}
Then
\begin{eqnarray}\label{le:eqn seven}	
X^{''}=H(\alpha ,\beta )=\begin{bmatrix}
1 & 1 & 1 & 1 & 1 & 1\\
1 & 1 & 1 &-1 & -1 & -1\\
1 & \o & \o^2 & \alpha  & \alpha \o  & \alpha \o^2 \\
1 & \o & \o^2 & -\alpha  & -\alpha \o & -\alpha \o^2 \\
1 & \o^2 & \o & \beta & \beta \o^2 & \beta \o \\
1 & \o^2 & \o & -\beta & -\beta \o^2 & -\beta \o
\end{bmatrix},
\end{eqnarray}
where $\o=e^{2\p i\over 3}$ and the parameters $\alpha$ and $\beta$ are both complex numbers of modulus one. So we complete this proof.
\end{proof}
\par
In fact, one can show that the two-paramrter family $H(\alpha ,\beta )$ in (\ref{le:eqn one}) is an affine family.

\begin{lemma}
\label{le:12 matrix elements}
Any $6\times 6$ CHM which contains no more than 12 imaginary elements is real equivalent to the following matrix $M_1$, or complex equivalent to the following symmetric matrix $M_2$.
\begin{eqnarray}\label{le:eqn eight}
M_1=\begin{bmatrix}
\o & \o  & 1 & 1 & 1 & 1\\
\o & -\o & -1 & 1 & -1 & 1\\
1 & 1 & \o & \o & 1 & 1\\
1 & -1 & -\o & \o & -1 & 1\\
1 & 1 & 1 & 1 & \o & \o\\
-1 & 1 & 1 & -1 & \o & -\o
\end{bmatrix}\ ,
M_2=
\begin{bmatrix}
i & 1 & 1 & 1 & 1 & 1\\
1 & i & 1 & 1 & -1 & -1\\
1 & 1 & i & -1 & 1 & -1\\
1 & 1 & -1 & i & -1 & 1\\
1 & -1 & 1 & -1 & i & 1\\
1 & -1 & -1 & 1 & 1 & i
\end{bmatrix}.
\end{eqnarray}
\end{lemma}

It's the main result of \cite{Liang2019}. One can show that if a CHM $H$ is real equivalent to $M_1$ then $H$ contains exactly four distinct elements $1,-1,\o$ and $-\o$. This property and Lemma \ref{le:12 matrix elements} will be used in the proof of Lemma \ref{le:three kinds of matrix elements and H2}. 
\par
The following lemma is about the CHM containing exactly two distinct elements. It is the first step in the studying 
of the CHM containing exactly specific distinct elements.

\begin{lemma}
\label{le:two elements}
The CHM containing exactly two distinct elements does not exist.
\end{lemma}
\begin{proof}
Suppose $B$ is a $6\times 6$ CHM which contains exactly two distinct elements. Up to complex equivalence, we assume that the distinct elements are $1$ and $s$. Obviously $s\neq -1$. Now we discuss this problem with two cases. 
\par
Case 1. $s$ is a pure imaginary number. Without loss of generality, we assume that $s=i$. Soppose there is a $2\times 6$ submatrix $B^{'}$ of $B$ such that there are two elements $i$ of $B^{'}$ on the same column. Then by the orthogonality of row vectors of $B^{'}$ we have 
\begin{eqnarray}\label{le:eqn nine}
1+x_1+x_2+x_3+x_4+x_5=0
\end{eqnarray}
where $x_k\in \{1,i,-i \}(k=1,2,...,5)$. One can show it is impossible. Similarly there is not a $6\times 2$ submatrix $B^{''}$ of $B$ such that there are two elements $i$ of $B^{''}$ on the same row. Hence the number of $i$ in $B$ is at most six. It makes $B$ no longer a CHM. So we have a contradiction. 
\par
Case 2. $s$ is not a pure imaginary number. We have $s+s^*\neq 0$. There is a diagonal unitary matrix 
\begin{eqnarray}
M=\rm{diag}(m_1,m_2,m_3,m_4,m_5,m_6)
\end{eqnarray}
such that $B_1=BM$ and elements of the first row of $B_1$ are one, where $m_k=1\ or\ s^*(k=1,2,...,6)$. So the elements of $B_1$ are $1,s$ or $s^*$. Suppose some row of $B_1$ other than the first row contains at least three $s$. By the orthogonality of this row vector and the first row vector of $B_1$ we have 
\begin{eqnarray}
s+s+s+y_1+y_2+y_3=0 \Longrightarrow 1+1+1+s^*y_1+s^*y_2+s^*y_3=0,
\end{eqnarray}
where $y_1,y_2,y_3\in \{1,s,s^* \}$. It lead to $s^*y_1=s^*y_2=s^*y_3=-1$, and it is a contradiction by $s$ is not a pure imaginary number. So any row of $B_1$ other than the first row has at most two $s$. 
Similarly we can show that any row of $B_1$ other than the first row has at most two $s^*$ or $1$. 
\par
So any row of $B_1$ other than the first row contains exactly two $s$, two $s^*$ and two $1$. 
Up to equivalent, we can assume that
\begin{eqnarray}\label{le:eqn ten}
B_1=\begin{bmatrix}
1 & 1 & 1 & 1 & 1 & 1\\
1 & 1 & s & s & s^* & s^*\\
a_0 & a_1 & a_2 & a_3 & a_4 & a_5 \\
b_0 & b_1 & b_2 & b_3 & b_4 & b_5 \\
c_0 & c_1 & c_2 & c_3 & c_4 & c_5 \\
d_0 & d_1 & d_2 & d_3 & d_4 & d_5
\end{bmatrix},
\end{eqnarray}
where $t_m\in \{1,s,s^* \}(m=0,1,...,5,\ t=a,b,c,d)$.
If the first column of $B_1$ contains $s^*$, then $m_1=s^*$. Other elements of the first column of $B_1$ are $s^*$ or $1$. Similarly if the first column of $B_1$ contains $s$, then $m_1=1$. So any column of $B_1$ does not contain $s$ and $s^*$ at the same time. Hence there is a monomial unitary matrix $P$ s.t. 
\begin{eqnarray}\label{le:eqn ten}
B_1P=\begin{bmatrix}
1 & 1 & 1 & 1 & s & s\\
1 & 1 & s & s & 1 & 1\\
a_0 & a_1 & a_2 & a_3 & sa_4 & sa_5 \\
b_0 & b_1 & b_2 & b_3 & sb_4 & sb_5 \\
c_0 & c_1 & c_2 & c_3 & sc_4 & sc_5 \\
d_0 & d_1 & d_2 & d_3 & sd_4 & sd_5
\end{bmatrix}
\end{eqnarray}
where $t_m\in \{1,s,s^* \}(m=0,1,\ t=a,b,c,d)$, $t_n\in \{1,s \}(n=2,3,\ t=a,b,c,d)$, $st_n\in \{1,s \}(n=4,5,\ t=a,b,c,d)$\par
If $a_2=a_3$, then by the orthogonality of the third and fourth column vectors of $B_1P$ we have 
\begin{eqnarray}
0=1+ss^*+a_2a_3^*+b_2b_3^*+c_2c_3^*+d_2d_3^*=1+1+1+b_2b_3^*+c_2c_3^*+d_2d_3^*,
\end{eqnarray}
where $b_2,b_3,c_2,c_3,d_2,d_3$ are $1$ or $s$. It leads to $b_2b_3^*=c_2c_3^*=d_2d_3^*=-1$, and it is a contradiction by $s$ is not a pure imaginary number. Hence $a_2\neq a_3$. Similarly we have $sa_4\neq sa_5$.
Up to equivalence, we assume that $a_2=sa_4=s,a_3=sa_5=1$. By the orthogonality of the first three row vectors of $B_1P$, we obtain that 
\begin{eqnarray}
\left\{
\begin{array}{lr}
1+1+s+s+s^*+s^*=0  \\
a_0+a_1+1+s^*+1+s=0 
\end{array}
\right.
\end{eqnarray}
One can work out $a_0=s,a_1=s^*$ or  $a_0=s^*,a_1=s$. 
Similarly we have
$$b_0=s,b_1=s^*\  or \  b_0=s^*,b_1=s,$$
$$c_0=s,c_1=s^*\  or \  c_0=s^*,c_1=s,$$
$$d_0=s,d_1=s^*\  or \  d_0=s^*,d_1=s.$$
Up to equivalence, we assume that $a_0=s,a_1=s^*$. So we have $b_0=s,b_1=s^*$, $c_0=s,c_1=s^*$, $d_0=s,d_1=s^*$.
Then by the orthogonality of the first and second column vectors of $B_1P$ we have 
$1+1+s(s^*)^*+s(s^*)^*+s(s^*)^*+s(s^*)^*=0$. Namely $2+4s^2=0$. Hence $|s|\neq 1$. It is a contradiction. So we complete this proof.
\end{proof}

\subsection{$H_2$-reducible matrices}
\label{subsec:h2 and mub}

In this section we review Theorem 11 of the paper \cite{karlsson11}. It characterizes a special family of CHMs, namely the $H_2$-reducible matrices. 
\begin{definition}
An $H_2$-reducible matrix is defined as a $6\times 6$ CHM  containing a $2\times 2$ Hadamard submatrix.
\end{definition}
\begin{lemma}\label{le:h2matrix}
The $H_2$-reducible CHM is equivalent to the CHM $H$ in \cite[Theorem 11]{karlsson11}, namely
\bea
\label{eq:h2-1}
H=&& 
\bma
F_2 & Z_1 & Z_2\\
Z_3 & {1\over2}Z_3AZ_1 & {1\over2}Z_3BZ_2\\
Z_4 & {1\over2}Z_4BZ_1 & {1\over2}Z_4AZ_2\\
\ema
\eea
where 
\begin{eqnarray}
&&
F_2=\bma1&1\\1&-1 \ema,	
\quad
Z_1=\bma1&1\\z_1&-z_1 \ema,	
\quad
Z_2=\bma1&1\\z_2&-z_2 \ema,
\\&&	
Z_3=\bma1&z_3\\1&-z_3 \ema,	
\quad
Z_4=\bma1&z_4\\1&-z_4 \ema,	
\\&&
A=
\bma 
A_{11} & A_{12}
\\
A_{12}^* & -A_{11}^*
\ema,
\quad
B=
\bma 
-1-A_{11} & -1-A_{12}
\\
-1-A_{12}^* & 1+A_{11}^*
\ema,
\end{eqnarray}
and
\begin{eqnarray}
&&
A_{11}=-{1\over2}+i{\sqrt3\over2}(\cos\t+e^{-i\phi}\sin\t),
\\&&	
A_{12}=-{1\over2}+i{\sqrt3\over2}(-\cos\t+e^{i\phi}\sin\t),
\\&&
\t,\phi\in[0,\pi),
\quad
\abs{z_j}=1.
\end{eqnarray}
\qed
\end{lemma}

\section{$H_2$-reducible matrices with three distinct elements }
\label{sec:H2}
In this section, we investigate the $H_2$-reducible matrices with three distinct elements. For convenience, let $a_i$ be the number of imaginary elements in the $i^{'}$th row of a CHM $M$. We refer to the array $im(M)$ $:=[a_1,a_2,a_3,a_4,a_5,a_6]$ as the imaginary array of $M$. It was first used in paper \cite{lma2020} and will simplify our proof.  As a result, the sum of $a_i(i=1,2,3,4,5,6)$ is the number of imaginary elements of $M$. 
We start the proof of Theorem \ref{le:three kinds of matrix elements and H2} with a preliminary lemma.
\begin{lemma}
\label{le:evenx}
If an $H_2$-reducible matrix $H$ contains exactly three distinct matrix elements $1$, $-1$ and $x$, then the numbers of $x$ of all rows of $H$ are all odd or $H$ is complex equivalent to the $M_2$ in (\ref{le:eqn eight}).
\end{lemma}

\begin{proof}
Suppose $H$ contains a $2\times 6$ submatrix $S$ which contains odd $x$. According to the orthogonality of row vectors of $S$, we obtain that $ax+bx^*+c=0$ where $a+b$ is odd and $a,b,c\in \mathbb{Z}$. Because $x$ is an imaginary number and $a+b$ is odd, $ax+bx^*$ is an imaginary number. We have $ax+bx^*+c\neq 0$, that is a contradiction. Hence all $2\times 6$ submatrices of $H$ contains even $x$. It leads to that the numbers of $x$ of all rows of $H$ are all odd or are all even.
\par
Let $im(H)=[a_1,a_2,a_3,a_4,a_5,a_6]$, where $H$ is an $H_2$-reducible matrix containing exactly three distinct matrix elements $1,-1$ and $x$. Using Lemma \ref{le:h2matrix}, $H$ contains nine $2\times 2$ Hadamard submatrices. Obviously every one of the submatrices contains no more than two $x$. Then the number of $x$ of $H$ is  at most $9\times 2=18$. That is $a_1+a_2+a_3+a_4+a_5+a_6\leq 18$.
\par
Suppose $a_1,a_2,...,a_6$ are all even, we know that $a_n\in \{0,2,4,6\}(n=1,2,...,6)$. If $a_m\geq 4$ for some $m\in \{1,2,...,6\}$, then let all elements of the $m^{'}$th row of $H$ multiply by $x^*$. We get the new matrix $H^{'}$ which is complex equivalent to $H$, and $im(H^{'})\leq 2\times 6=12$. 
\par
When $x\neq \pm i$, $H^{'}$ might contain $1,-1,x,x^*$ or $-x^*$. Lemma \ref{le:12 matrix elements} shows that $H^{'}$ is real equivalent to $M_1$ in (\ref{le:eqn eight}) or complex equivalent to $M_2$ in (\ref{le:eqn eight}). If $H^{'}$ is real equivalent to $M_1$, then $H^{'}$ contains exactly four distinct elements $1,-1,\o$ and $-\o$. Because $x\neq \pm i$, $x\neq -x^*$ and $x\neq -(-x^*)$. Hence $H^{'}$ contains exactly four distinct elements $1,-1,x^*$ and $-x^*$. One can show that $H$ contains exactly four distinct elements $1,-1,x$ and $-x$. However $H$ contains exactly three distinct matrix elements $1,-1$ and $x$. That is a contradiction. So $H^{'}$ is complex equivalent to $M_2$ in (\ref{le:eqn eight}) and so is $H$.
\par
When $x= \pm i$, $H^{'}$ might contain $1,-1,x, x^*$. Lemma \ref{le:12 matrix elements} shows that $H^{'}$ is real equivalent to $M_1$ in (\ref{le:eqn eight}) or complex equivalent to $M_2$ in (\ref{le:eqn eight}). Because $x= \pm i$, $H^{'}$ is not real equivalent to $M_1$ in (\ref{le:eqn eight}). Hence $H$ is complex equivalent to the $M_2$ in (\ref{le:eqn eight}). 
\par 
So we complete this proof.
\end{proof}

\begin{theorem}
\label{le:three kinds of matrix elements and H2}
If an $H_2$-reducible matrix $H$ contains exactly three distinct matrix elements, then $H$ is complex equivalent to the $M_2$ in (\ref{le:eqn eight}).
\end{theorem}

\begin{proof}
Suppose the three distinct matrix elements of $H$ are $1$,$k$ and $t$. There are two diagonal unitary matrices $D_1,D_2$ such that all elements of the first row and column of $D_1HD_2$ are one. The $5\times 5$ submatrix of the lower right corner of $D_1HD_2$ contains 7 distinct  matrix elements, they are $1,k,t,k^*,t^*,kt^*,tk^*$. Because $H$ is an $H_2$-reducible matrix, we obtain $-1\in \{1,k,t,k^*,t^*,kt^*,tk^*\}$. If $tk^*=-1$, then $k^*H$ contains exactly three distinct matrix elements 1,$-1$ and $k^*$. If $k^*=-1$, then $k=k^*=-1$. Up to equivalence, one can get similar conclusion for $kt^*=-1$ and $t^*=-1$. So we can assume that $H$ contains exactly three distinct matrix elements $1,-1$ and $k$. \par
Now we proceed with two cases in terms of whether $k=\pm i$.\par
Case 1. $k\neq \pm i$. 
\par 
Let $im(H)=[a_1,a_2,a_3,a_4,a_5,a_6]$, where $H$ is an $H_2$-reducible matrix containing exactly three distinct matrix elements $1,-1$ and $k$. Using Lemma \ref{le:h2matrix}, $H$ contains nine $2\times 2$ Hadamard submatrices. Obviously every one of these submatrices contains no more than two $k$. Then the number of $k$ of $H$ is  at most $9\times 2=18$. That is $a_1+a_2+a_3+a_4+a_5+a_6\leq 18$. Using Lemma \ref{le:evenx}, $a_1,a_2,...,a_6$ are all odd or $H$ is complex equivalent to the $M_2$ in (\ref{le:eqn eight}). 
\par
Suppose $a_1,a_2,...,a_6$ are all odd, we know that $a_n\in \{1,3,5\}(n=1,2,...,6)$. 

If $H$ contains the following submatrices
\begin{eqnarray}\label{le:kkk1}
\begin{bmatrix}
k & k & k & d_1 & d_2 & d_3\\
k & k & k & d_4 & d_5 & d_6
\end{bmatrix},
\end{eqnarray}
where $d_n(n=1,2,...,6)$ are 1 or $-1$. Then using Lemma \ref{le:rank one} we obtain that $H$ contains more than three elements. It is a contradiction. 

If $H$ contains the following submatrices
\begin{eqnarray}\label{le:kkk2}
\begin{bmatrix}
k & k & k & d_1 & d_2 & d_3\\
d_4 & d_5 & d_6 & k & k & k
\end{bmatrix},
\end{eqnarray}
where $d_n(n=1,2,...,6)$ are 1 or $-1$. Then by the orthogonality we have $k^*(d_4+d_5+d_6)+k(d_1+d_2+d_3)^*=0$. Because $k=\pm i$ and $d_n(n=1,2,...,6)$ are 1 or $-1$, one can show that $k^*(d_4+d_5+d_6)+k(d_1+d_2+d_3)^*\neq 0$. It is a contradiction. 

If $H$ contains the following submatrices
\begin{eqnarray}\label{le:kkk3}
\begin{bmatrix}
k & k & k & d_1 & d_2 & d_3\\
d_4 & k & k & k & d_5 & d_6
\end{bmatrix},
\end{eqnarray}
where $d_n(n=1,2,...,6)$ are 1 or $-1$. Then by the orthogonality we have $k^*d_4+2+kd_1^*+d_2^*d_5+d_3^*d_6=0$. Because $k=\pm i$ and $d_n(n=1,2,...,6)$ are 1 or $-1$, one can show that $k^*d_4+2+kd_1^*+d_2^*d_5+d_3^*d_6\neq 0$. It is a contradiction. 

Suppose the number of 3 in $\{a_1,a_2,a_3,a_4,a_5,a_6\}$ is more than 4. Because $H$ does not contain these submatrices in (\ref{le:kkk1}), (\ref{le:kkk2}) and (\ref{le:kkk3}),
$H$ contains the submatrix
\begin{eqnarray}\label{le:kkk}
\begin{bmatrix}
k & k & k & d_1 & d_2 & d_3\\
d_4 & d_5 & k & k & k & d_6\\
d_7 & k & d_8 & k & d_9 & k\\
k & d_{10} & d_{11} & d_{12} & k & k
\end{bmatrix}
\end{eqnarray}
up to complex equvalence, where $d_n(n=1,2,...,12)$ are 1 or -1. Then by the pigeonhole principle, the remaining row of $H$ containing three $k$ leads to a contradiction with (\ref{le:kkk1}), (\ref{le:kkk2}) and (\ref{le:kkk3}). Hence the number of 3 in $\{a_1,a_2,a_3,a_4,a_5,a_6\}$ is at most 4. 

Suppose the number of 3 in $\{a_1,a_2,a_3,a_4,a_5,a_6\}$ is less than 4. If $a_n\geq 5$ for some $n$, let all elements of the $n^{'}$th row of $H$ multiply by $k^*$. We get the new matrix $H^{''}$ which is complex equivalent to $H$, and $im(H^{''})\leq 3\times 3+1+1+1=12$. Lemma \ref{le:12 matrix elements} shows that $H$ is complex equivalent to $M_2$ in (\ref{le:eqn eight}).\par
Hence the only unsloved case is that the number of 3 in $\{a_1,a_2,a_3,a_4,a_5,a_6\}$ is four. One can show that $H$
contains the submatrix in (\ref{le:kkk}) up to complex equvalence.  

Suppose $5\in \{a_1,a_2,a_3,a_4,a_5,a_6\}$, By the pigeonhole principle and (\ref{le:kkk}) we know that there is a row of $H$ containing five $k$ and $kk^*+kk^*+kk^*+k(d_x+d_y)^*+rd_z^*=0$, where $d_x,d_y,d_z,r$ are 1 or -1. By $k\neq \pm i$, one can show that $kk^*+kk^*+kk^*+k(d_x+d_y)^*+rd_z^*=3+k(d_x+d_y)^*+rd_z^^*\neq 0$.  It is a contradiction. 

Suppose $1\in \{a_1,a_2,a_3,a_4,a_5,a_6\}$, By the pigeonhole principle and (\ref{le:kkk}) we know that there is a row of $H$ containing one $k$ and $k(d_x+d_y+d_z)^*+k^*d_u+r_1^*d_v+r_2^*d_w=0$, where $d_x,d_y,d_z.d_u,d_v,d_w,r_1,r_2$ are 1 or -1. By $k\neq \pm i$, one can show that $k(d_x+d_y+d_z)^*+k^*d_u+r_1^*d_v+r_2^*d_w\neq 0$.  It is a contradiction. 

Hence the number of 3 in $\{a_1,a_2,a_3,a_4,a_5,a_6\}$ is not four and we complete this proof of Case 1.

Case 2. $k=\pm i$. Without loss of generality we assume $k=i$.

There is a diagonal unitary matrix $D$ such that all elements of the first row of $HD$ are one. So all elements of $HD$ are 1, $-1$, $i$ and $-i$. Let $im(HD)=[0,a_1,a_2,a_3,a_4,a_5]$, one can show that $a_n(n=1,2,...,5)\neq 1,3,5$, so $a_n(n=1,2,...,5)\in \{0,2,4\}$. If $a_n=4$ for some $n$, let all elements of the $n^{'}$th row of $HD$ multiply by $i$. Then we get a new matrix $H_i$ and $im(HD)=[0,d_1,d_2,d_3,d_4,d_5]$, where $d_n(n=1,2,...,5)\in \{0,2\}$. Hence $0+d_1+d_2+d_3+d_4+d_5\leq 0+2\times 5=10<12$. Lemma \ref{le:12 matrix elements} shows that $H$ is complex equivalent to $M_2$ in (\ref{le:eqn eight}).\par
By Case 1 and Case 2 we complete this proof.

\end{proof}
Lemma \ref{le:three kinds of matrix elements and H2} is the first main result of this paper. It shows that an $H_2$-reducible matrix containing exactly three distinct elements is complex equivalent to the $M_2$ in (\ref{le:eqn eight}). In fact, we hope to find all of the CHMs which contain exactly three distinct elements. We conjecture the CHMs with exactly three distinct elements are complex equivalent to $M_2$ in (\ref{le:eqn eight}) or the Tao matrix.\tbc
\section{CHMs containg a $3\times 3$ Hadamard submatrix}
\label{sec:CHMs 33}
In this section, we investigate the CHM which contains a $3\times 3$ Hadamard submatrix. First we introduce a useful lemma.

\begin{lemma}
\label{le:fro}
If $S$ is a CHM containing exactly three distinct elements and the first row of $S$ are one. Then $S$ is complex equivalent to the Tao matrix.
\end{lemma}

\begin{proof}
Suppose three distinct elements of $S$ are $1,a,b$, and
\begin{eqnarray}\label{le:eqn 11}
S=\begin{bmatrix}
1 & 1 & 1 & 1 & 1 & 1\\
a_0 & a_1 & a_2 & a_3 & a_4 & a_5 \\
b_0 & b_1 & b_2 & b_3 & b_4 & b_5 \\
c_0 & c_1 & c_2 & c_3 & c_4 & c_5 \\
d_0 & d_1 & d_2 & d_3 & d_4 & d_5 \\
f_0 & f_1 & f_2 & f_3 & f_4 & f_5
\end{bmatrix}
\end{eqnarray}
where all elements of $S$ are complex numbers of modulus one.
If $a_0=a_1=a_2=a$, then $a_3=a_4=a_5=-a$. Hence by the orthogonality of the first three row vectors of $S$ we have 
\begin{eqnarray}
\left\{
\begin{array}{lr}
b_0+b_1+b_2+b_3+b_4+b_5=0  \\
a^*(b_0+b_1+b_2)-a^*(b_3+b_4+b_5)=0 
\end{array}
\right.
\end{eqnarray}
So $b_0+b_1+b_2=0$. Using Lemma \ref{le:abcd} $(i)$, $b_i\neq b_j$ and $b_i+b_j\neq 0$ $(i,j=1,2,3,i\neq j)$.
So there are at least four numbers in $\{b_0,b_1,b_2,a,-a\}$ which are different from each other. This is a contradiction. Similarly any row of $S$ other than the first row not contains more than three $1$ or $b$. Hence any row of $S$ other than the first row contains exactly two $1$, two $a$ and two $b$. Then by orthogonality of the first two row vectors of $S$ we have $1+1+a+a+b+b=0$. Using Lemma \ref{le:abcd} ($i$), $a=\o, b=\o^2$ or $a=\o^2, b=\o$ where $\o=e^{2\p i\over 3}$. Hence $S$ is complex equivalent to 
\begin{eqnarray}\label{le:eqn 13}
S^{'}=\begin{bmatrix}
1 & 1 & 1 & 1 & 1 & 1\\
1 & 1 & \o & \o & \o^2 & \o^2 \\
1 & x_1 & 1 & x_2 & x_3 & x_4 \\
1 & y_1 & y_2 & 1 & y_3 & y_4 \\
1 & z_1 & z_2 & z_3 & 1 & z_4 \\
1 & u_1 & u_2 & u_3 & u_4 & 1
\end{bmatrix}
\end{eqnarray}
where $v_k\in \{\o , \o^2 \}(v=x,y,z,u,\ k=1,2,...,4)$. Any row of $S^{'}$ other than the first row contains exactly two $1$, two $\o$ and two $\o^2 $. Up to complex equivalence, we have the following steps
$$\begin{bmatrix}
1 & 1 & 1 & 1 & 1 & 1\\
1 & 1 & \o & \o & \o^2 & \o^2 \\
1 & x_1 & 1 & x_2 & x_3 & x_4 \\
1 & y_1 & y_2 & 1 & y_3 & y_4 \\
1 & z_1 & z_2 & z_3 & 1 & z_4 \\
1 & u_1 & u_2 & u_3 & u_4 & 1
\end{bmatrix}$$
$$\Downarrow (x_3\neq x_4)$$
$$\begin{bmatrix}
1 & 1 & 1 & 1 & 1 & 1\\
1 & 1 & \o & \o & \o^2 & \o^2 \\
1 & x_1 & 1 & x_2 & \o & \o^2 \\
1 & y_1 & y_2 & 1 & y_3 & y_4 \\
1 & z_1 & z_2 & z_3 & 1 & z_4 \\
1 & u_1 & u_2 & u_3 & u_4 & 1
\end{bmatrix}$$
$$\Downarrow (Lemma \ \ref{le:rank one})$$
$$\begin{bmatrix}
1 & 1 & 1 & 1 & 1 & 1\\
1 & 1 & \o & \o & \o^2 & \o^2 \\
1 & x_1 & 1 & x_2 & \o & \o^2 \\
1 & y_1 & y_2 & 1 & \o^2 & \o \\
1 & z_1 & z_2 & z_3 & 1 & z_4 \\
1 & u_1 & u_2 & u_3 & u_4 & 1
\end{bmatrix}$$
$$\Downarrow (Lemma \ \ref{le:rank one})$$
$$\begin{bmatrix}
1 & 1 & 1 & 1 & 1 & 1\\
1 & 1 & \o & \o & \o^2 & \o^2 \\
1 & \o & 1 & \o^2 & \o & \o^2 \\
1 & y_1 & y_2 & 1 & \o^2 & \o \\
1 & z_1 & z_2 & z_3 & 1 & z_4 \\
1 & u_1 & u_2 & u_3 & u_4 & 1
\end{bmatrix}$$
$$\Downarrow (Lemma \ \ref{le:rank one})$$
$$\begin{bmatrix}
1 & 1 & 1 & 1 & 1 & 1\\
1 & 1 & \o & \o & \o^2 & \o^2 \\
1 & \o & 1 & \o^2 & \o & \o^2 \\
1 & \o & \o^2 & 1 & \o^2 & \o \\
1 & z_1 & z_2 & z_3 & 1 & z_4 \\
1 & u_1 & u_2 & u_3 & u_4 & 1
\end{bmatrix}$$
$$\Downarrow (u_2\neq u_3,\ z_2\neq z_3)$$
$$\begin{bmatrix}
1 & 1 & 1 & 1 & 1 & 1\\
1 & 1 & \o & \o & \o^2 & \o^2 \\
1 & \o & 1 & \o^2 & \o & \o^2 \\
1 & \o & \o^2 & 1 & \o^2 & \o \\
1 & z_1 & \o^2 & \o & 1 & z_4 \\
1 & u_1 & \o & \o^2 & u_4 & 1
\end{bmatrix}\ or \ \begin{bmatrix}
1 & 1 & 1 & 1 & 1 & 1\\
1 & 1 & \o & \o & \o^2 & \o^2 \\
1 & \o & 1 & \o^2 & \o & \o^2 \\
1 & \o & \o^2 & 1 & \o^2 & \o \\
1 & z_1 & \o & \o^2 & 1 & z_4 \\
1 & u_1 & \o^2 & \o & u_4 & 1
\end{bmatrix}$$
$$\Downarrow (Lemma \ \ref{le:rank one})$$
$$\begin{bmatrix}
1 & 1 & 1 & 1 & 1 & 1\\
1 & 1 & \o & \o & \o^2 & \o^2 \\
1 & \o & 1 & \o^2 & \o & \o^2 \\
1 & \o & \o^2 & 1 & \o^2 & \o \\
1 & \o^2 & \o^2 & \o & 1 & \o \\
1 & \o^2 & \o & \o^2 & \o & 1
\end{bmatrix}\ or \ \begin{bmatrix}
1 & 1 & 1 & 1 & 1 & 1\\
1 & 1 & \o & \o & \o^2 & \o^2 \\
1 & \o & 1 & \o^2 & \o & \o^2 \\
1 & \o & \o^2 & 1 & \o^2 & \o \\
1 & \o^2 & \o & \o^2 & 1 & \o \\
1 & \o^2 & \o^2 & \o & \o & 1
\end{bmatrix}$$
$$\Downarrow (Lemma\  \ref{le:rank one})$$
$$S^{''}=\begin{bmatrix}
1 & 1 & 1 & 1 & 1 & 1\\
1 & 1 & \o & \o & \o^2 & \o^2 \\
1 & \o & 1 & \o^2 & \o & \o^2 \\
1 & \o & \o^2 & 1 & \o^2 & \o \\
1 & \o^2 & \o & \o^2 & 1 & \o \\
1 & \o^2 & \o^2 & \o & \o & 1
\end{bmatrix}$$
It is easy to check that $S^{''}$ is complex eqiuvalent to Tao matrix. So we complete this proof.
\end{proof}
Then we introduce a inference of Lemma \ref{le:fro}.
\begin{lemma}\label{le:inference}
If $M_1$ is a CHM containing exactly three distinct elements $1,\o ,\o^2 $ where $\o=e^{2\p i\over 3}$. Then $M_1$ is complex equivalent to the Tao matrix.
\end{lemma}
\begin{proof}
There is a CHM $M_2$ s.t. $M_2$ is complex equivalent to $M_1$ and the first row of $M_2$ are one. It is easy to obtain that $M_2$ also contains exactly three distinct matrix elements $1,\o ,\o^2 $ where $\o=e^{2\p i\over 3}$. Using Lemma \ref{le:fro} we complete this inference.
\end{proof}

\begin{lemma}
\label{le:33matrix}
Any CHM containing a $3\times 3$ Hadamard submatrix is complex equivalent to the Tao matrix or the matrix from the two-parameter family $H(\alpha , \beta)$ in Lemma \ref{le:rank one}.
\end{lemma}

\begin{proof}
Suppose $U$ is a $3\times 3$ Hadamard matrix, and $U$ is a submatrix of a CHM $H$. $U$ is complex equivalent to $U_0=\begin{bmatrix}
1 & 1 & 1 \\
1 & \o & \o^2 \\
1 & \o^2 & \o 
\end{bmatrix}$ where $\o=e^{2\p i\over 3}$. By Lemma \ref{le:abcd} ($i$) we have $H$ is complex equivalent to 
\begin{eqnarray}
A_1=\begin{bmatrix}
1 & 1 & 1 & 1 & 1 & 1\\
1 & \o & \o^2 & a & a\o & a\o^2\\
1 & \o^2 & \o & b & b\o^2 & b\o\\
1 & c & d & x_1y_1 & x_2y_1 & x_3y_1\\
1 & c\o & d\o^2 & x_1y_2 & x_2y_2\o & x_3y_2\o^2\\
1 & c\o^2 & d\o & x_1y_3 & x_2y_3\o^2 & x_3y_3\o
\end{bmatrix}
\end{eqnarray}
or 
\begin{eqnarray}
A_2=\begin{bmatrix}
1 & 1 & 1 & 1 & 1 & 1\\
1 & \o & \o^2 & a & a\o & a\o^2\\
1 & \o^2 & \o & b & b\o^2 & b\o\\
1 & c & d & x_1y_1 & x_2y_1 & x_3y_1\\
1 & c\o & d\o^2 & x_1y_2 & x_2y_2\o^2 & x_3y_2\o\\
1 & c\o^2 & d\o & x_1y_3 & x_2y_3\o & x_3y_3\o^2
\end{bmatrix}
\end{eqnarray}where $\o=e^{2\p i\over 3}$ and $a,b,c,d,s_n(s=x,y,z,\ n=1,2,3)$ are complex numbers of modulus one.
\par
Case 1. $H$ is complex equivalent to $A_1$.
By the orthogonality of all columns of $A_1$, we have
\begin{eqnarray}\label{le:eqn 29}
\left\{
\begin{aligned}
1+a+b+x_1(y_1+y_2+y_3)& = & 0 \\
1+a+b+c^*x_2(y_1+y_2+y_3)& = & 0\\
1+a+b+d^*x_3(y_1+y_2+y_3)& = & 0
\end{aligned}
\right.
\end{eqnarray} and 
\begin{eqnarray}
\left\{
\begin{aligned}
1+c+d+y_1(x_1+x_2+x_3)& = & 0 \\
1+c+d+a^*y_2(x_1+x_2+x_3)& = & 0\\
1+c+d+b^*y_3(x_1+x_2+x_3)& = & 0
\end{aligned}
\right.
\end{eqnarray}
If $y_1+y_2+y_3=0$, then we have $1+a+b=0$. Using Lemma \ref{le:abcd} ($i$) we know that $a=\o,b=\o^2$ or $a=\o^2,b=\o$. Using Lemma \ref{le:rank one}, $H$ is complex equivalent to the matrix from the two-parameter family $H(\alpha , \beta)$ in Lemma \ref{le:rank one}. So we only consider $1+a+b\neq 0$. By (\ref{le:eqn 29}) we know $x_1=c^*x_2=d^*x_3$. Similarly we have $y_1=a^*y_2=b^*y_3$. Hence 
\begin{eqnarray}
A_1=\begin{bmatrix}
1 & 1 & 1 & 1 & 1 & 1\\
1 & \o & \o^2 & a & a\o & a\o^2\\
1 & \o^2 & \o & b & b\o^2 & b\o\\
1 & c & d & x_1y_1 & cx_1y_1 & dx_1y_1\\
1 & c\o & d\o^2 & ax_1y_1 & acx_1y_1\o & adx_1y_1\o^2\\
1 & c\o^2 & d\o & bx_1y_1 & bcx_1y_1\o^2 & bdx_1y_1\o
\end{bmatrix}.
\end{eqnarray} From the first column and the fourth column we have $(1+a+b)(1+x_1y_1)=0$. So $x_1y_1=-1$, and \begin{eqnarray}
A_1=\begin{bmatrix}
1 & 1 & 1 & 1 & 1 & 1\\
1 & \o & \o^2 & a & a\o & a\o^2\\
1 & \o^2 & \o & b & b\o^2 & b\o\\
1 & c & d & -1 & -c & -d\\
1 & c\o & d\o^2 & -a & -ac\o & -ad\o^2\\
1 & c\o^2 & d\o & -b & -bc\o^2 & -bd\o
\end{bmatrix}.
\end{eqnarray}From the first column and the fifth column we have $(1+a\o+b\o^2)(1-c)=0$. If $1+a\o+b\o^2=0$, then using Lemma \ref{le:abcd} ($i$) we know that $a=1,b=1$ or $a=\o , b=\o^2$. Using Lemma \ref{le:rank one}, $H$ is complex equivalent to the matrix from the two-parameter family $H(\alpha , \beta)$ in Lemma \ref{le:rank one}. So we only consider $1+a\o+b\o^2\neq 0$ and it means $c=1$. Similarly we have $d=a=b=1$. Then $H$ is complex equivalent to
\begin{eqnarray}
\begin{bmatrix}
1 & 1 & 1 & 1 & 1 & 1\\
1 & \o & \o^2 & 1 & \o & \o^2\\
1 & \o^2 & \o & 1 & \o^2 & \o\\
1 & 1 & 1 & -1 & -1 & -1\\
1 & \o & \o^2 & -1 & -\o & -\o^2\\
1 & \o^2 & \o & -1 & -\o^2 & -\o
\end{bmatrix}
\end{eqnarray} and it is from the two-parameter family $H(\alpha , \beta)$ in Lemma \ref{le:rank one}.
\par
Case 2. $H$ is complex equivalent to $A_2$. By the orthogonality of all columns of $A_2$, we have
\begin{eqnarray}\label{le:eqn 34}
\left\{
\begin{aligned}
1+a+b+x_1(y_1+y_2+y_3)& = & 0 \\
1+a+b+c^*x_2(y_1+y_2\o+y_3\o^2)& = & 0\\
1+a+b+d^*x_3(y_1+y_2\o^2+y_3\o)& = & 0
\end{aligned}
\right.
\end{eqnarray}
\begin{eqnarray}\label{le:eqn 35}
\left\{
\begin{aligned}
1+a\o+b\o^2+x_2(y_1+y_2w^2+y_3w)& = & 0 \\
1+a\o+b\o^2+d^*x_1(y_1+y_2w+y_3w^2)& = & 0\\
1+a\o+b\o^2+c^*x_3(y_1+y_2+y_3)& = & 0
\end{aligned}
\right.
\end{eqnarray}
\begin{eqnarray}\label{le:eqn 36}
\left\{
\begin{aligned}
1+a\o^2+b\o+x_3(y_1+y_2\o+y_3\o^2)& = & 0 \\
1+a\o^2+b\o+c^*x_1(y_1+y_2\o^2+y_3\o)& = & 0\\
1+a\o^2+b\o+d^*x_2(y_1+y_2+y_3)& = & 0
\end{aligned}
\right.
\end{eqnarray}
If $1+a+b=0$, using Lemma \ref{le:abcd} ($i$) we know that $a=\o,b=\o^2$ or $a=\o^2,b=\o$. Using Lemma \ref{le:rank one}, $H$ is complex equivalent to the matrix from the two-parameter family $H(\alpha , \beta)$ in Lemma \ref{le:rank one}. 
Similarly we only consider \begin{eqnarray}\label{le:eqn 37}
\left\{
\begin{aligned}
1+a+b \neq 0 \\
1+a\o^2+b\o \neq 0\\
1+a\o+b\o^2 \neq 0
\end{aligned}
\right.
\end{eqnarray}From (\ref{le:eqn 34})(\ref{le:eqn 35})(\ref{le:eqn 36}) and (\ref{le:eqn 37}) we obtain that
\begin{eqnarray}
\left\{
\begin{aligned}
x_1(y_1+y_2+y_3) &=& c^*x_2(y_1+y_2\o+y_3\o^2)&= & d^*x_3(y_1+y_2\o^2+y_3\o) \\
c^*x_3(y_1+y_2+y_3) &=& d^*x_1(y_1+y_2\o+y_3\o^2)&= & x_2(y_1+y_2\o^2+y_3\o)\\
d^*x_2(y_1+y_2+y_3)&=& x_3(y_1+y_2\o+y_3\o^2)&= & c^*x_1(y_1+y_2\o^2+y_3\o)
\end{aligned}
\right.
\end{eqnarray}Then we could work out 
\begin{eqnarray}
\left\{
\begin{aligned}\label{le:eqn 39}
c^3 &=& d^3&= & 1 \\
x_1^3 &=& x_2^3&= & x_3^3
\end{aligned}
\right.
\end{eqnarray}
Similarly we have 
\begin{eqnarray}
\left\{
\begin{aligned}\label{le:eqn 40}
a^3 &=& b^3&= & 1 \\
y_1^3 &=& y_2^3&= & y_3^3
\end{aligned}
\right.
\end{eqnarray}
Next we analysis the equation $c^3=d^3=1$ by cases.
\par
Subcase 1. $\{c=1,d=1\}$, $\{c=\o ,d=\o^2\}$ or $\{c=\o^2 ,d=\o \}$. At this time there is a $2\times 3$ submatrix with rank one of $A_2$. Using Lemma \ref{le:rank one}, $H$ is complex equivalent to the matrix from the two-parameter family $H(\alpha , \beta)$ in Lemma \ref{le:rank one}. 
\par
Subcase 2. $c=d=\o$. Then $1+c\o +d\o^2=1+c\o^2+d\o $. By the orthogonality of rows of $A_2$, we have
\begin{eqnarray}\label{le:eqn 41}
\left\{
\begin{aligned}
1+c+d+a^*y_2(x_1+x_2\o+x_3\o^2)& = & 0 \\
1+c+d+b^*y_3(x_1+x_2\o^2+x_3\o)& = & 0\\
1+c+d+y_1(x_1+x_2+x_3)& = & 0
\end{aligned}
\right.
\end{eqnarray}
\begin{eqnarray}\label{le:eqn 42}
\left\{
\begin{aligned}
1+c\o^2+d\o+a^*y_1(x_1+x_2\o^2+x_3\o)& = & 0 \\
1+c\o^2+d\o+b^*y_2(x_1+x_2+x_3)& = & 0\\
1+c\o^2+d\o+y_3(x_1+x_2\o+x_3\o^2)& = & 0
\end{aligned}
\right.
\end{eqnarray}
\begin{eqnarray}\label{le:eqn 43}
\left\{
\begin{aligned}
1+c\o+d\o^2+a^*y_3(x_1+x_2+x_3)& = & 0 \\
1+c\o+d\o^2+b^*y_1(x_1+x_2\o+x_3\o^2)& = & 0\\
1+c\o+d\o^2+y_2(x_1+x_2\o^2+x_3\o)& = & 0
\end{aligned}
\right.
\end{eqnarray}
If $1+c+d=0$, using Lemma \ref{le:abcd} ($i$) we know that $c=\o,d=\o^2$ or $c=\o^2,d=\o$. Using Lemma \ref{le:rank one}, $H$ is complex equivalent to the matrix from the two-parameter family $H(\alpha , \beta)$ in Lemma \ref{le:rank one}. 
Similarly we only consider \begin{eqnarray}\label{le:eqn 44}
\left\{
\begin{aligned}
1+c+d \neq 0 \\
1+c\o^2+d\o \neq 0\\
1+c\o+d\o^2 \neq 0
\end{aligned}
\right.
\end{eqnarray}
From (\ref{le:eqn 42})(\ref{le:eqn 43}) and (\ref{le:eqn 44}) we have
\begin{eqnarray}\label{le:eqn 45}
\left\{
\begin{aligned}
a^*y_3(x_1+x_2+x_3)=b^*y_2(x_1+x_2+x_3)\\
b^*y_1(x_1+x_2\o+x_3\o^2)=y_3(x_1+x_2\o+x_3\o^2)\\
y_2(x_1+x_2\o^2+x_3\o)=a^*y_1(x_1+x_2\o^2+x_3\o) 
\end{aligned}
\right.
\end{eqnarray}
By (\ref{le:eqn 41})(\ref{le:eqn 44}) and (\ref{le:eqn 45}) we have 
\begin{eqnarray}\label{le:eqn 46}
y_1=by_3=ay_2.
\end{eqnarray}
Hence from (\ref{le:eqn 34}) and (\ref{le:eqn 46}) we obtain that 
\begin{eqnarray}\label{le:eqn 47}
1+a+b+x_1y_1(1+a^*+b^*)=0.
\end{eqnarray}
By (\ref{le:eqn 34})(\ref{le:eqn 47}) and Lemma \ref{le:abcd} ($iii$) we have $x_1y_1\in \{1,\o ,\o^2 ,-1,-\o ,-\o^2 \}$.\par
If $x_1y_1\in \{-1,-\o ,-\o^2 \}$, then by (\ref{le:eqn 47}) we have
\begin{eqnarray}\label{le:eqn 48}
-x_1y_1=\dfrac{1+a+b}{1+a^*+b^*}\in \{1,\o ,\o^2 \}.
\end{eqnarray}By (\ref{le:eqn 40}) and (\ref{le:eqn 37}) we have 
\par
If $a=b=\o $, then $\dfrac{1+a+b}{1+a^*+b^*}=\dfrac{1+2\o }{1+2\o^2 }\notin \{1,\o ,\o^2 \}$,\par
If $a=b=\o^2$, then $\dfrac{1+a+b}{1+a^*+b^*}=\dfrac{1+2\o^2 }{1+2\o}\notin \{1,\o ,\o^2 \}$,\par
If $a=1,b=\o$ or $a=\o,b=1$, then $\dfrac{1+a+b}{1+a^*+b^*}=\dfrac{2+\o }{2+\o^2}\notin \{1,\o ,\o^2 \}$,\par
If $a=1,b=\o^2$ or $a=\o^2,b=1$, then $\dfrac{1+a+b}{1+a^*+b^*}=\dfrac{2+\o^2 }{2+\o}\notin \{1,\o ,\o^2 \}$.\par
Hence $x_1y_1\in \{1,\o ,\o^2 \}$. Similarly we have $x_2y_1\in \{1,\o ,\o^2 \}$ and $x_3y_1\in \{1,\o ,\o^2 \}$. So 
all elements of $A_2$ are 1, $\o$ or $\o^2$. Using Lemma \ref{le:fro}, $A_2$ is complex equivalent to the Tao matrix.
\par
Subcase 3. Using the method of Subcase 2, we could show that whatever $c,d$ from (\ref{le:eqn 39}) and (\ref{le:eqn 44}) are $A_2$ is complex equivalent to Tao matrix or the matrix from the two-parameter family $H(\alpha , \beta)$ in Lemma \ref{le:rank one}.\par
From now we have proven this proof. 
\end{proof}
The content of this section is the uppermost result in this paper. We have proven that the CHMs containing a $3\times 3$ Hadamard submatrix is complex equivalent to the Tao matrix or the matrix from the two-parameter family $H(\alpha , \beta)$ in Lemma \ref{le:rank one}. It might be a new way to think about the Tao matrix.

\section{Conclusions}
\label{sec:con}
In this paper, we investigate a series of special CHMs and we have a specific classification for these CHMs. The next target for us is to find more non-$H_2$-reducible matrices like the Tao matrix. We believe more non-$H_2$-reducible matrices will lead to the solution of the famous MUB problem in dimension six.

\bibliographystyle{unsrt}

\bibliography{mengfan=6x6real20}

\end{document}